\documentclass[aps,prd,floats,floatfix,twocolumn,showpacs,superscriptaddress,nofootinbib]{revtex4}
\usepackage{amsmath,amsfonts}
\usepackage{graphicx}

\def\prl{Phys. Rev. Lett.}
\def\prd{Phys. Rev. D}

\def\lrr{Living Rev. Relat.}

\def\jmp{J. Math. Phys.}

\begin{document}

\title{Shells around black holes: the effect of freely specifiable  
quantities in Einstein's constraint equations}

\author{Keith Matera}

\affiliation{Department of Physics and Astronomy, Bowdoin College,  
Brunswick, ME 04011, USA}

\author{Thomas W. Baumgarte}

\altaffiliation{Also at Department of Physics, University of  
Illinois, Urbana, IL 61801}
\affiliation{Department of Physics and Astronomy, Bowdoin College,  
Brunswick, ME 04011, USA}

\author{Eric Gourgoulhon}

\affiliation{Laboratoire Univers et Th\'eories (LUTH), Observatoire  
de Paris, CNRS, Universit\'e Paris Diderot; 5 place Jules Janssen,  
92190 Meudon, France}

\date{8 November 2007}

\begin{abstract}
We solve Einstein's constraint equations in the conformal thin-sandwich decomposition to model thin shells of non-interacting  
particles in circular orbit about a non-rotating black hole.  We use  
these simple models to explore the effects of some of the freely  
specifiable quantities in this decomposition on the physical content  
of the solutions.  Specifically, we adopt either maximal slicing or  
Kerr-Schild slicing, and make different choices for the value of the  
lapse on the black hole horizon.  For one particular choice of these  
quantities the resulting equations can be solved analytically; for  
all others we construct numerical solutions.  We find that these  
different choices have no effect on our solutions when they are  
expressed in terms of gauge-invariant quantities.
\end{abstract}

\maketitle

\section{Introduction}

A 3+1 decomposition of Einstein's equations results in a set of  
constraint equations, which constrain the gravitational fields at all  
instants of coordinate time, and a set of evolution equations, which  
propagate the fields forward in time (e.g.~\cite 
{Dar27,ArnDM62,Yor79}).  The four constraint equations can constrain  
only a subset of the gravitational fields.  Therefore, the constraint  
equations can be solved, for example for the construction of initial  
data, only after the constrained variables have been separated from  
freely specifiable ones, and after suitable choices have been made  
for the latter (see e.g. \cite{Pfe04,Gou07b} for reviews).

The constrained variables are separated from the freely specifiable  
ones by choosing a decomposition of the constraint equations.  The  
conformal thin-sandwich decomposition \cite{Yor99,PfeY03} has been  
particularly popular for the construction of quasiequilibrium data;  
it has been used extensively, for example, to model compact binaries  
containing black holes or neutron stars (see, e.g., \cite 
{Coo00,BauS03} for reviews).  In the conformal thin-sandwich  
formalism, the spatial metric is conformally decomposed into a  
conformal factor and the conformally related metric, and the  
extrinsic curvature into its trace and a traceless part.  In the 
so-called extended version \cite{PfeY03}, the freely specifiable  
variables are the conformally related metric and the trace of the  
extrinsic curvature together with their time derivatives (which we  
may set to zero to construct equilibrium data), and the constrained  
variables are the lapse, the shift, and the conformal factor.

Black holes may be constructed within the conformal thin-sandwich  
formalism by excising the black hole interior, and imposing suitable  
inner boundary conditions.  In particular, these boundary conditions  
may be chosen so that the black hole is momentarily isolated, or in  
equilibrium (see \cite{Coo02,CooP04,JarGM04}, also compare the  
isolated horizon formalism laid out in \cite{AshK04,DreKSS03,GouJ06}  
and references therein).  As discussed in detail in \cite{CooP04},  
these geometric conditions lead to boundary conditions on some of the  
constrained variables in the thin-sandwich formalism, namely the  
conformal factor and the shift vector.  The boundary condition for   
the lapse, however, remains arbitrary.

Some of the choices in this formalism will clearly have an effect on  
the physical content of the solution.   We can expect to find  
equilibrium solutions only if we set the time derivatives of the  
conformally related metric and the trace of the extrinsic curvature  
to zero.  Also, a conformally flat solution is physically distinct  
from solutions that are not conformally flat.  The choice of the  
trace of the extrinsic curvature, or mean curvature, is usually  
associated with an initial temporal gauge, and the lapse plays a  
similar role.  It is less clear, then, whether or how the mean  
curvature and the boundary condition on the lapse affect the solutions.

In \cite{CooP04}, the authors found that sequences of binary black  
holes, and in particular their innermost stable circular orbit, do  
depend on the horizon lapse for their example of a non-maximal  
slicing, i.e.~non-zero mean curvature.  This finding, however, may be  
an artifact of their particular choice of the mean curvature, namely  
a superposition of two copies of its analytical value for a single  
Schwarzschild black hole expressed in Kerr-Schild coordinates (see  
(\ref{K_KS}) below), one for each companion in the binary.  As the  
authors caution, the resulting background geometry then depends on  
binary separation, making the physical meaning of these sequences  
somewhat arguable.

We consider a very simple physical system in order to analyze  
whether, at least in this context, the choice of the mean curvature  
and the horizon lapse affect the physical content of the solutions.   
Specifically, we solve the constraint equations in the thin-sandwich  
decomposition to construct thin shells of non-interacting, isotropic  
particles in circular orbit about a Schwarzschild black hole (compare  
\cite{SkoB02}, whose results we generalize to account for the black  
hole).  For one particular choice of the mean curvature and the  
horizon lapse the equations can be solved analytically (see Appendix  
\ref{appA}), and we construct numerical solutions for many others.   
We find that these different choices have no effect on our solutions  
when they are expressed in terms of gauge-invariant quantities.

\section{Basic equations}

\subsection{Constraint equations}

We write the spacetime metric $g_{ab}$ in the form
\begin{equation}
g_{ab} dx^a dx^b = - \alpha^2 dt^2 + \gamma_{ij} (dx^i + \beta^i dt) 
(dx^j + \beta^j dt),
\end{equation}
where $\alpha$ is the lapse function, $\beta^i$ the shift vector, and  
$\gamma_{ij}$ the spatial metric.  We further decompose the latter as
\begin{equation}\label{metric}
\gamma_{ij} =  \psi^4 \bar \gamma_{ij},
\end{equation}
where $\psi$ is a conformal factor and $\bar \gamma_{ij}$ a  
conformally related metric.  We then solve Einstein's constraint  
equations in the conformal thin-sandwich decomposition (see \cite 
{Yor99,PfeY03}, as well as \cite{Coo00,BauS03,Gou07b} for reviews).    
Specifically, the Hamiltonian constraint becomes
\begin{eqnarray}\label{psiConstraint}
\bar D^2 \psi = \frac{1}{8} \psi \bar{R} +  \frac{1}{12}\psi^5 K^2 - 
\frac{1}{8} \psi^{-7} \bar A_{ij} \bar A^{ij} - 2 \pi \psi^5 \rho_N.
\end{eqnarray}
Here $\rho_N = T_{ab} n^a n^b $ is the energy density as measured by  
a normal observer, $\bar D^2 = \bar \gamma^{ij} \bar D_i \bar D_j$,  
and $\bar D_i$ and $\bar{R}$ are the covariant derivative and the  
Ricci scalar associated with the metric $\bar \gamma_{ij}$.  We have  
also split the extrinsic curvature $K_{ij}$ into its trace $K$ and a  
traceless part $A_{ij}$ according to
\begin{equation}
K_{ij} =A_{ij} +\frac{1}{3} \gamma_{ij}  K =\psi^{-2} \bar A_{ij} +  
\frac{1}{3} \gamma_{ij} K.
\end{equation}
 From the evolution equation for the spatial metric we can express $ 
\bar A^{ij}$ as
\begin{equation}\label{matrixA}
\bar A^{ij} = \frac{1}{2\bar \alpha} \left( \left( \bar L \beta  
\right)^{ij}-\bar u^{ij} \right).
\end{equation}
Here $\bar \alpha = \psi^{-6} \alpha$ and $\bar{u}^{ij} = \partial_t  
\bar \gamma_{ij}$, and the
conformal Killing operator  $\bar{L}$ is defined as
\begin{equation}\label{Lbar}
\left( \bar L \beta \right) ^{ij} \equiv \bar D^i \beta^j + \bar D^j  
\beta^i -\frac{2}{3} \bar \gamma^{ij} \bar D_k \beta^k .
\end{equation}
The momentum constraint can now be written as
\begin{eqnarray}\label{betaConstraint}
\left( \bar \Delta_L \beta \right)^i &=& \left( \bar L \beta \right)^ 
{ij} \bar D_j \ln \left( \bar \alpha \right)+
\bar \alpha \bar D_j \left( \bar \alpha^{-1} \bar u^{ij} \right)  
\nonumber \\
&&+\frac{4}{3} \bar \alpha \psi^6 \bar D^i K + 16 \pi \bar \alpha  
\psi^{10} j^i,
\end{eqnarray}
where $(\bar \Delta_L \beta)^i = D_j (\bar L \beta)^{ij}$ is a vector  
Laplacian, and $j^i = - \gamma^{ia} n^b T_{ab}$ is the mass current  
as measured by a normal observer.  Finally, the trace of the  
evolution equation for $K_{ij}$, combined with the Hamiltonian  
constraint, results in
\begin{eqnarray}\label{alphaConstraint}
\bar{D}^2 \left( \alpha \psi \right) & = & \alpha \psi \Big( \frac{7} 
{8} \psi^{-8} \bar A_{ij} \bar A^{ij} + \frac{5}{12} \psi^4 K^2 +  
\frac{1}{8}\bar{R} \\
&&+  2 \pi \psi^4 \left( \rho + 2S \right) \Big) - \psi^5 \partial_t  
K + \psi^5 \beta^i \bar{D}_i K, \nonumber
\end{eqnarray}
where $S=\gamma^{ij} T_{ij}$ the trace of the spatial stress.

The above equations form a set of equations for the lapse $\alpha$,  
the shift $\beta^i$ and the conformal factor $\psi$.  Before these  
equations can be solved, however, we have to make choices for the  
freely specifiable quantities $\bar \gamma_{ij}$, $\bar u_{ij} =  
\partial_t \bar \gamma_{ij}$, $K$ and $\partial_t K$.  For the  
construction of quasiequilibrium data it is natural to choose $\bar u_ 
{ij} =0$ and $\partial_t K = 0$.  We will also restrict our analysis  
to spherical symmetry, where we may assume conformal flatness, $\bar  
\gamma_{ij} = \eta_{ij}$, without loss of generality.  Here $\eta_{ij} 
$ is the flat metric in whatever coordinate system.  We will,  
however, experiment with different choices for $K$ (see equations  
(\ref{K}) below), as well as with different boundary conditions for  
the lapse $\alpha$ (see (\ref{alphaHorBound}) below).

With these choices, and in spherical symmetry, the above equations  
simplify dramatically.  We write the spatial metric as
\begin{equation}
\gamma_{ij} dx^i dx^j  =  \psi^4 \left( dr^2 + r^2 (d\theta^2 +  
\sin^2 \theta d\phi^2) \right),
\end{equation}
where $r$ is the isotropic radial coordinate.
The shift $\beta^i$ is now purely radial, and we abbreviate $\beta  
\equiv \beta^r$.  We may evaluate (\ref{Lbar}) to find
\begin{equation}
\left(  \bar L \beta \right)^{ij} =
\displaystyle - \frac{2}{3r}
\left(
\begin{array}{ccc}
-2 r^2 & 0 & 0 \\
0 &  1 & 0 \\
0 & 0 & \displaystyle \frac{1}{\sin^2\theta} \\
\end{array}
\right)
\partial_r \left(\dfrac{\beta}{r}\right),
\end{equation}
so that $\bar{A}_{ij}\bar{A}^{ij}$ becomes
\begin{equation}
\bar{A}_{ij}\bar{A}^{ij} = \frac{2}{3 \bar{\alpha}^2}r^2\left 
(\partial_r \frac{\beta}{r}\right)^2.
\end{equation}
The Hamiltonian constraint (\ref{psiConstraint}) can then be written as
\begin{subequations} \label{DiffEq}
\begin{eqnarray}\label{psiConstraintMod}
r \partial_r^2 \psi & + & 2 \partial_r \psi + r \frac{\psi^5}{12}  
\left( \frac{ \left(
\partial_r \beta -\beta / r \right)^2}{\alpha^2} - K^2 \right)  
\nonumber \\
& = &  -2 \pi \psi^5 \rho_N,
\end{eqnarray}
the momentum constraint (\ref{betaConstraint}) as
\begin{eqnarray}\label{betaConstraintMod}
\partial_r^2 \beta & + & \left( \displaystyle \frac{2}{r} - \frac 
{\partial_r \alpha}{\alpha} + 6 \frac{\partial_r \psi}{\psi} \right)  
\left(\partial_r \beta  - \displaystyle \frac{\beta}{r}\right)   
\nonumber \\
& = & \alpha \partial_r K + 12 \pi \psi^4 \alpha j^i,
\end{eqnarray}
and the lapse equation (\ref{alphaConstraint}) as
\begin{eqnarray}\label{alphaConstraintMod}
\partial_r^2 \left( \alpha \psi \right) &=& \alpha\psi \Big(\frac{7  
\psi^4}{12\alpha^2} \big( \partial_r \beta
  - \frac{\beta}{r} \big)^2 + \frac{5}{12} \psi^4 K^2 \nonumber \\
&& + 2 \pi \psi^4 \left(\rho + 2 S \right) + \psi^5 \beta \partial_r  
K \Big).
\end{eqnarray}
\end{subequations}
Here we have expressed all quantities in terms of those variables  
that are used in our code.

In the above equations the trace of the extrinsic curvature $K$ can  
still be chosen arbitrarily.  Following \cite{CooP04} we consider two  
different possibilities, namely maximal slicing
\begin{subequations} \label{K}
\begin{equation}
K_{\rm MS} = 0
\end{equation}
and Kerr-Schild slicing
\begin{equation} \label{K_KS}
K_{\rm KS} = \frac{2 M_{\rm BH}}{R^2} \left(1+\frac{2M_{\rm BH}}{R} 
\right)^{-3/2} \left(1+\frac{3M_{\rm BH}}{R} \right).
\end{equation}
\end{subequations}
Here $R$ is the areal radius, $R=\psi^2 r$, and we identify $M_{\rm  
BH}$ with the black hole's irreducible mass (see (\ref{Mirr}) below).  Kerr-Schild
coordinates are identical to ingoing Eddington-Finkelstein coordinates.

Throughout this paper, we use the subscript $\rm_{BH}$ to refer to  
the black hole, and $\rm_{SH}$ to
indicate a property of the shell.

\subsection{Boundary conditions}
\label{sec:BoundCond}

At spatial infinity we impose asymptotic flatness, which results in  
the boundary conditions
\begin{equation}\label{infBound}
\psi \rightarrow 1,  ~~~~
\alpha \rightarrow 1, ~~~~
\beta \rightarrow 0
\end{equation}
as $r \rightarrow \infty$.

We excise the black hole interior inside an isotropic radius $r_{\rm  
BH}$ and impose the black hole equilibrium boundary conditions of  
\citet{CooP04} on the resulting excision surface ${\mathcal S}$  
(compare the notion of isolated horizons laid out in \cite 
{AshK04,GouJ06}).  In particular, the condition
\begin{equation}\label{psiHorBound}
\left. m^{ab} \nabla_a k_b \right|_{\mathcal S}  =
\left. m^{ij} \left( D_i s_j - K_{ij} \right) \right|_{\mathcal S} = 0,
\end{equation}
where $s^i$ is the outward-pointing unit normal to the horizon,  
ensures that this surface corresponds to a marginally trapped surface  
(apparent horizon), while the condition
\begin{equation}\label{betaHorBound}
\left. \beta_{\perp} \right|_{\mathcal S}  \equiv \left. \beta^r s_r  
\right|_{\mathcal S} =  \left. \alpha \right|_{\mathcal S}
\end{equation}
ensures that the coordinate system tracks the horizon.  (The  
tangential components of the shift vanish identically in spherical  
symmetry.)  In our case, (\ref{psiHorBound}) becomes
\begin{subequations} \label{InBoundCond}
\begin{equation} \label{psiHorBoundMod}
\left( \partial_r \psi + \frac{\psi}{2r} + \frac{\psi^3}{6}  \left 
(\frac{\partial_r \beta - \beta/r}{\alpha} - K  \right) \right)_ 
{\mathcal S} = 0
\end{equation}
and (\ref{betaHorBound})
\begin{equation} \label{betaHorBoundMod}
\left. \beta \psi^2 \right|_{\mathcal S} = \left. \alpha \right|_ 
{\mathcal S}.
\end{equation}

As discussed in \cite{CooP04}, the boundary condition for the lapse $ 
\alpha$ is arbitrary.  We will experiment with the Dirichlet boundary  
condition
\begin{equation}\label{alphaHorBound}
\left. \alpha \right|_{\mathcal S} = \alpha_{\rm AH},
\end{equation}
\end{subequations}
and will compare results for values of $\alpha_{\rm AH}$ ranging from  
zero to unity in increments of 0.1.

\subsection{Matter equations}

We consider a spherically symmetric shell of isotropic, non-interacting 
particles in circular orbit about the black hole (compare  
\cite{SkoB02}).
The rest energy (baryon mass) of a spherically symmetric matter  
source may be written as
\begin{equation}
M_{\rm SH} = \int \rho_0 u^t \sqrt{-g}d^3x
        = 4 \pi \int \rho_0 W \psi^6 r^2 dr,
\end{equation}
where $W \equiv - n_{\alpha}u^{\alpha} = \alpha u^t$ is the Lorentz  
factor between a normal observer $n^a$ and an observer comoving with  
the fluid $u^a$.   For an infinitesimally thin shell we can then  
identify the rest energy density (baryon density) as
\begin{equation}\label{restEnergyDensity}
\rho_0 = \frac{M_{\rm SH}}{4 \pi W \psi^6 r^2 } \delta( r-r_{\rm SH}),
\end{equation}
where $r_{\rm SH}$ is the (isotropic) radius of the shell.

Since the particles are non-interacting, their stress-energy tensor  
is that of a pressureless
fluid (dust): $T_{ab} = \rho_0 u_a u_b$.  The matter sources $\rho_N 
$, $S$ and $j^r$ in equations (\ref{psiConstraintMod}),  (\ref 
{betaConstraintMod}) and  (\ref{alphaConstraintMod}) can therefore be  
expressed as
\begin{eqnarray}
\rho_N & = & \rho_0 W^2, \nonumber \\
S & = & \rho_0 (W^2 - 1), \\
j^r & = & \rho_0 \beta W^2 / \alpha. \nonumber
\end{eqnarray}
The delta function in these matter sources leads to a discontinuity  
in the derivatives of the solutions.
We can find the jump in these derivatives by integrating equations  
(\ref{betaConstraint}), (\ref{psiConstraint}) and (\ref 
{alphaConstraint}) from $r_{\rm SH} - \epsilon$ to $r_{\rm SH} +  
\epsilon$, which, in the limit $\epsilon \rightarrow 0$, results in  
the jump conditions
\begin{eqnarray}\label{jumpConditions}
\partial_r \beta_+ - \partial_r \beta_- &=& \frac{3WM_{\rm SH}\beta} 
{r_{\rm SH}^2 \psi^2}, \nonumber \\
\partial_r \psi_+ - \partial_r \psi_- &=& \frac{-M_{\rm SH}W}{2r_{\rm  
SH}^2 \psi},  \\
\partial_r \left( \alpha \psi \right)_+ - \partial_r \left( \alpha  
\psi \right)_- &=&
\alpha \psi \left( \frac{M_{\rm SH} \left(3 W^2-2 \right)}{2r_{\rm SH} 
^2 W \psi^2}\right). \nonumber
\end{eqnarray}

Since the particles are non-interacting, their 4-velocity $u^a$ obeys  
the geodesic equation
\begin{equation}
du^a/d\tau+\Gamma^a_{bc}u^bu^c=0,
\end{equation}
Assuming circular orbits with $u^r = 0$ and $d u^r/d\tau = 0$ and,  
without loss of generality, focussing on a particle in the equatorial  
plane, we find
\begin{equation}\label{geodesicMod}
\Gamma^r_{tt}\left(u^t\right)^2+\Gamma^r_{\phi\phi}\left(u^{\phi} 
\right)^2=0.
\end{equation}
Using the normalization condition $u_a u^a=-1$, which is equivalent to 
$1=W^2- \gamma_{ij} u^i u^j$, we evaluate (\ref{geodesicMod})
to find the geodesic condition
\begin{equation}\label{geodesicCondition}
1-W^{-2}=
\left( \frac{\beta^2}{r^2} - \dfrac{\psi^3\beta\left(\psi\partial_r 
\beta+2\beta\partial_r\psi\right)-\alpha\partial_r\alpha}
{\left(2r\partial_r\psi+\psi\right)\psi^3r} \right) \frac{ \psi^4 r^2} 
{\alpha^2}.
\end{equation}
for circular orbits.  To compute the Christoffel symbols we averaged  
the derivatives of the gravitational field variables inside and  
outside the shell (compare \cite{SkoB02}).

\subsection{Diagnostics}

We compute the ADM and Komar masses using the expressions
(see e.g. Chap.~7 of \cite{Gou07a})
\begin{equation}
M_{\rm ADM} = -\frac{1}{2\pi} \int_{\partial\Sigma} \bar D_i \psi \,  
d \bar S^i
\end{equation}
and
\begin{equation}
M_{\rm K} = \frac{1}{4\pi} \int_{\partial \Sigma}  \left( D_i \alpha  
- \beta^j K_{ij} \right) dS^i \, ,
\end{equation}
where $dS^i$ is the outward pointing unit surface element of a closed  
surface at infinity.  In our case, these expressions reduce to
\begin{equation}\label{ADM}
M_{\rm ADM} = \lim_{r\rightarrow\infty} -2 r^2 \partial_r \psi
\end{equation}
and
\begin{equation}\label{Komar}
M_{\rm K} = \lim_{r\rightarrow\infty} r^2 \partial_r \alpha \, .
\end{equation}
For all configurations considered in this paper, the two mass  
expressions are found
to be equivalent to within the accuracy of our code when the geodesic  
equation (\ref{geodesicCondition}) is used to force the particles  
into circular orbit (compare \cite{SkoB02}). This is in agreement  
with a general theorem about the equality
of ADM and Komar masses established by Beig \cite{Bei78} and
Ashtekar \& Magnon-Ashtekar \cite{AshM79}.
We also define the binding energy as
\begin{equation} \label{Mbind}
E_{\rm B} = \frac{M_{\rm ADM/K}}{M_{\rm BH} + M_{\rm SH}} - 1,
\end{equation}
either terms of the ADM or Komar mass.  Finally, we compute the black  
hole's irreducible mass from the area ${\mathcal A}$ of the apparent  
horizon,
\begin{equation} \label{Mirr}
M_{\rm irr} = \left( \frac{\mathcal A}{16 \pi} \right)^{1/2} = \frac 
{R}{2} = \frac{r_{\rm BH} \psi_{\rm BH}^2}{2}.
\end{equation}

\section{Numerics}

\subsection{Code}

We developed a pseudo-spectral code to solve the differential  
equations (\ref{DiffEq}) subject to the boundary conditions (\ref 
{infBound}) and (\ref{InBoundCond}) as well as the jump conditions  
(\ref{jumpConditions}), using Chebyshev polynomials as basis functions
(see \cite{GraN07} for a recent review of spectral methods).
Equations (\ref{betaConstraintMod}) and (\ref{alphaConstraintMod})  
can be solved directly, while equation (\ref{psiConstraintMod}) has  
to be linearized and then solved iteratively.

One complication arises as a consequence of the jump conditions (\ref 
{jumpConditions}).  Representing the solution functions across these  
jumps as a linear combination of the continuous Chebyshev polynomials  
would result in undesirable Gibbs phenomena.  To avoid this problem,  
we solve the equations in two separate domains inside and outside the  
shell, each one represented by $N$ Chebyshev polynomials.   The $N$  
coefficients can then be determined by evaluating the equation at $N$  
collocation points in each domain, and the jump conditions (\ref 
{jumpConditions}) can then be imposed exactly as matching conditions  
between the two sets of Chebyshev polynomials.

Each set of Chebyshev polynomials is $T_n(s)$ with $0\leq n\leq N-1$ and
$s\in[-1,1]$.  We map the inner region into the interval $[-1,1]$  
with the transformation
\begin{equation} \label{transI}
s_{\rm I}=\frac{2 r_{\rm BH}r_{\rm SH}/r-r_{\rm BH}-r_{\rm SH}}{r_ 
{\rm BH}-r_{\rm SH}}
\end{equation}
and the outer region with
\begin{equation} \label{transO}
s_{\rm O}=-2r_{\rm SH}/r +1.
\end{equation}
Our computational domain therefore extends to $r = \infty$, and we  
can evaluate the masses (\ref{ADM}) and (\ref{Komar}) exactly.

In addition to the choices for $\alpha_{\rm AH}$ and $K$, our  
solution depends on the parameters $M_{\rm SH}$ and $r_{\rm SH}$ in  
(\ref{restEnergyDensity}), and the excision radius $r_{\rm BH}$.  To  
construct a solution of given black hole mass $M_{\rm BH}$, we need  
to iterate over $r_{\rm BH}$ until the resulting irreducible black  
hole mass (\ref{Mirr}) agrees with the desired black hole mass $M_ 
{\rm BH}$ to within a certain pre-determined tolerance.  For a given  
shell radius $r_{\rm SH}$ a further iteration is needed to fix the  
Lorentz factor $W$ in such a way that the solution satisfies the  
geodesic condition (\ref{geodesicCondition}), and the particles are  
in circular orbit.  In practice, we instead fix $W$ and then iterate  
over $r_{\rm SH}$ until (\ref{geodesicCondition}) is satisfied.

For each case we start the iteration from the analytical solution for  
$K = 0$ and $\alpha_{\rm AH} = 0$ provided in Appendix \ref{appA},  
and continue until the solution has converged.  Specifically, our  
convergence criterion requires that the relative change between  
iteration steps in any of the fields is less then $10^{-10}$ at all  
collocation points.

\subsection{Tests}

We tested our program for a number of different known vacuum  
solutions expressing the Schwarzschild geometry in different  
coordinate systems, as well as an analytical solution describing thin  
shells around black holes that we derive in Appendix \ref{appA}.

As a first vacuum test (for which we set $M_{\rm SH} = 0$ and $r_{\rm  
SH} = 10M_{\rm BH}$), we considered Schwarzschild in ``standard"  
isotropic coordinates, representing the symmetry plane in a 
Carter-Penrose diagram.  In our code, we can produce the solution
\begin{equation} \label{SSiso}
\psi = 1 + \frac{M}{2r},~~~~ \alpha \psi = 1 - \frac{M}{2r},~~~~  
\beta =0
\end{equation}
by choosing $K = K_{\rm MS} = 0$ and $\alpha_{\rm AH} = 0$.  After  
applying the transformations (\ref{transI}) and (\ref{transO}), these  
solutions become linear in our code's coordinates $s_{\rm I}$ and $s_ 
{\rm O}$, meaning that the solution can be represented exactly in  
terms of the first two Chebyshev polynomials.  For any $N \geq 1$,  
our code therefore converges to the correct solution within the  
predetermined tolerances.

\begin{figure}[t]
\includegraphics[width=3in]{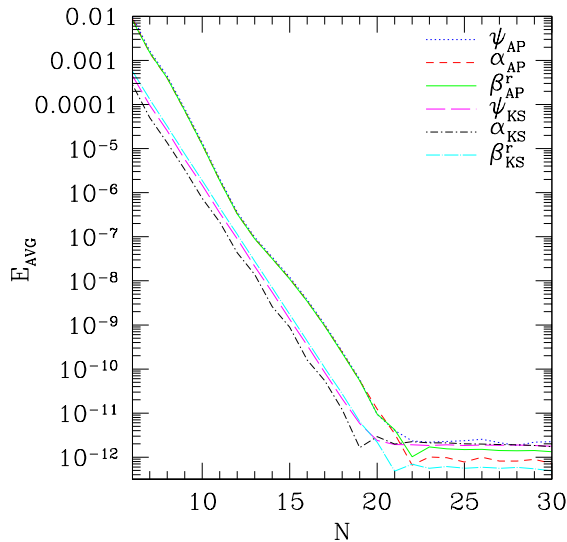}
\caption{The average error over all collocation points $E^N$ as a  
function of the number of collocation points $N$ for two analytic  
representations of the Schwarzschild geometry in isotropic  
coordinates.   The label $AP$ denotes the analytic puncture solution  
presented in \cite{BauN07}, while the label $KS$ denotes the Kerr- 
Schild solution in \cite{CooP04}.  The errors drop off exponentially  
as a function of the number of collocation points $N$, until a  
`floor' of specified tolerance has been reached.}
\label{APKSError}
\end{figure}

More interesting are the two isotropic representations of the  
Schwarzschild geometry presented in \cite{CooP04} and \cite{BauN07}.   
The former is a transformation of Kerr-Schild (Eddington-Finkelstein)  
coordinates to isotropic coordinates, keeping the same time slicing,  
which we can produce by choosing $K = K_{\rm KS}$ and $\alpha_{\rm  
AH} = 1/\sqrt{2}$ in our code.  The latter is an isotropic  
representation of a maximal slice (with the critical parameter $C = 3  
\sqrt{3}M^2/4$, see \cite{EstWCDST73,Rei73}), which has recently  
attracted interest as an analytic ``puncture" solution (compare \cite 
{HanHBGSO06}).   We can produce this solution by choosing $K = K_{\rm  
MS} = 0$ and $\alpha_{\rm AH} = 3\sqrt{3} / 16$ in our code.

To measure the deviation from an analytic solution, we compute the  
average of the absolute error $\epsilon_n$ at all $N$ collocation  
points in each of the two domains,
\begin{equation}
E^{N} = \frac{1}{2N} \sum_{n=1}^{2N} | \epsilon_n |.
\end{equation}
where the collocation points $N+1 \leq n \leq 2N$ are in the region  
outside of the shell.  In Fig.~\ref{APKSError} we graph this error as  
a function of the number of collocation points $N$.  As expected, the  
errors fall off exponentially for all variables, until they reach a  
floor corresponding to the predetermined tolerance.

\begin{figure}[t]
\includegraphics[width=3in]{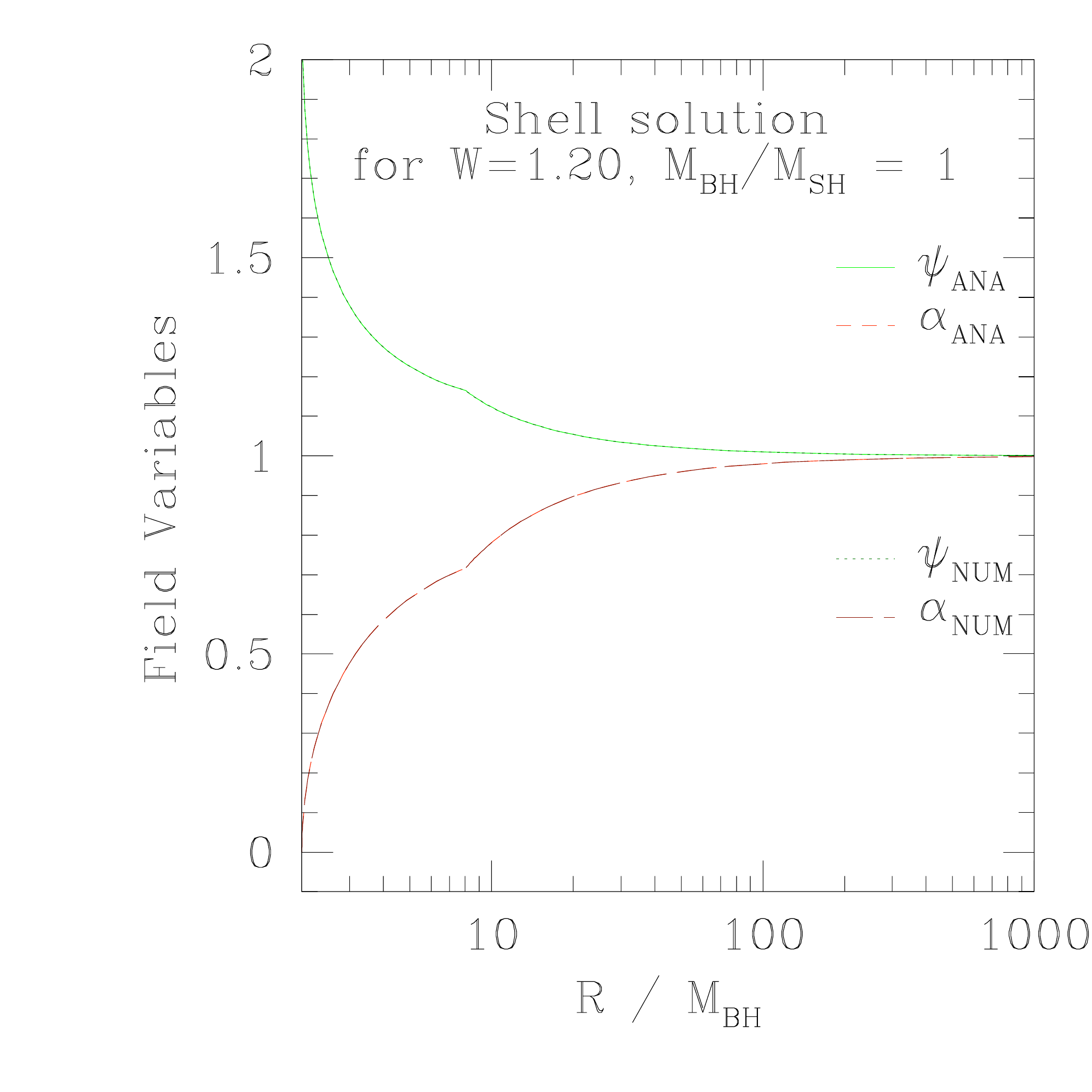}
\caption{The conformal factor $\psi$ and the lapse $\alpha$ as a function of areal  
radius for the analytic shell solution ($K = 0$ and $\alpha_{\rm EH}=0$) for a 
Lorentz factor of $W=1.20$ and a mass ratio of $M_{\rm BH}/M_{\rm SH}=1$, 
corresponding to an areal shell radius of about $8.013 M_{\rm BH}$.  The field variables
are continuous across the shell, but, according to the jump conditions (\ref{jumpConditions}) their derivatives are not.  Our numerical solutions agree with the analytical ones to within better 
than $10^{-10}$, making the error far smaller than the line width in the graph.}
\label{shellPlot}
\end{figure}

In addition to these vacuum solutions we also consider an analytic  
solution describing thin shells of non-interacting particles around a  
static black hole.
As we demonstrate in Appendix \ref{appA}, we can solve the  
differential equations (\ref{DiffEq}) subject to the boundary  
conditions (\ref{infBound}) and (\ref{InBoundCond}) as well as the  
jump conditions (\ref{jumpConditions}) analytically for maximal  
slicing and $\alpha_{\rm AH} = 0$.  As for the solution (\ref 
{SSiso}), the field variables become linear in our code's variables  
$s_{\rm I}$ and $s_{\rm O}$ (see (\ref{analyticPsiSH}) and (\ref 
{analyticAlphaSH}) below), so that they can be represented exactly  
for any $N \geq 1$.  In addition to testing the solution of the field  
equations, however, this test also verifies that our code correctly  
solves the jump conditions at the shell.  As in example, we show in Fig.~\ref{shellPlot} 
the analytic and numerical solutions for the lapse $\alpha$ and the conformal factor $\psi$ 
as a function of areal radius $R$ for a Lorentz factor of $W=1.20$ and a mass ratio 
$M_{BH}/M_{SH}=1$.  Our numerical solutions agree with the analytical ones to within better
than $10^{-10}$, making them indistinguishable in the plot.

\section{Results}

\begin{figure}[t]
\includegraphics[width=3in]{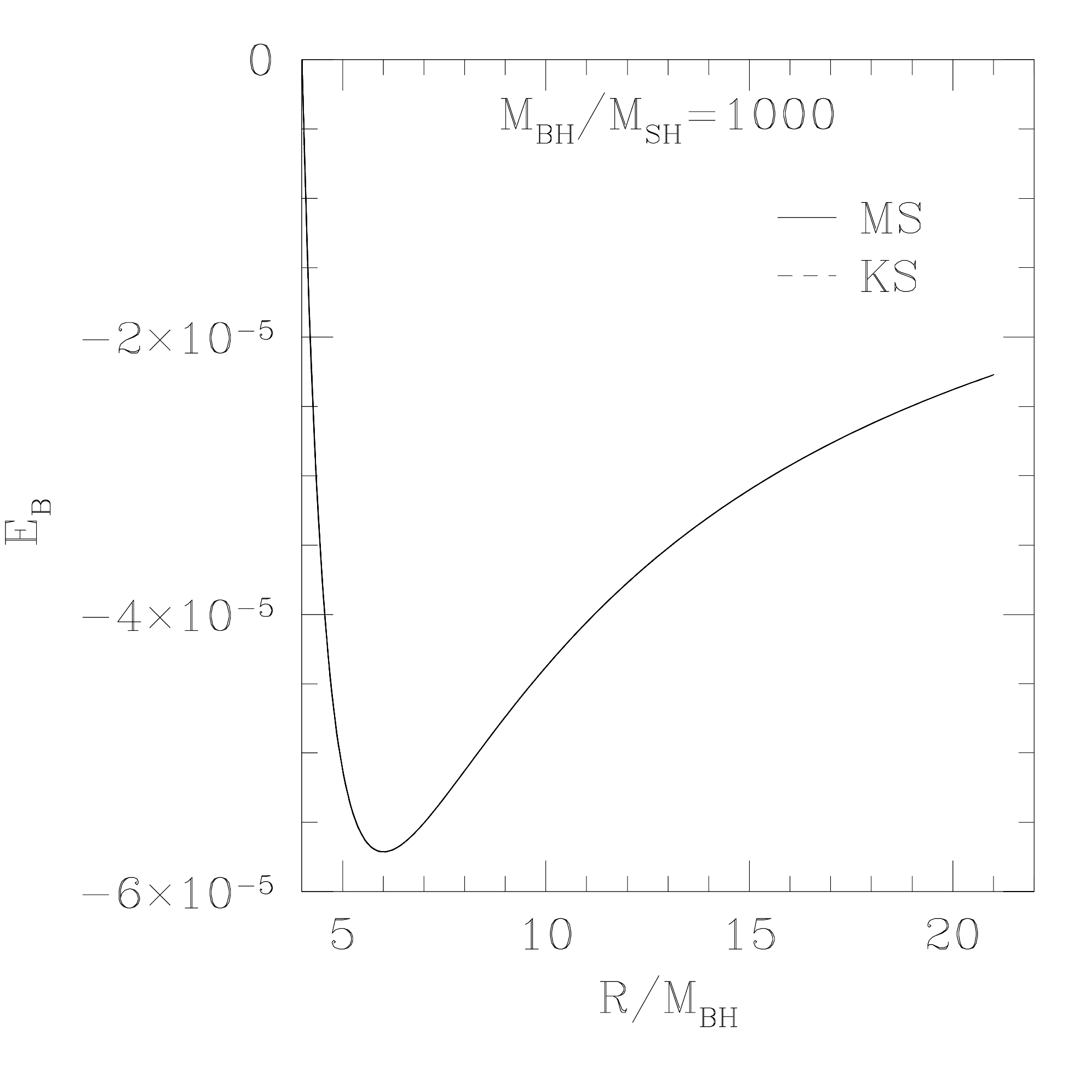}
\caption{The ADM binding energy (\ref{Mbind}) $E_{\rm B}$ as a  
function of the shell's areal radius $R$ for an extreme-mass-ratio  
sequence with $M_{BH}/M_{SH} = 1000$.  Values of the horizon lapse $ 
\alpha_{\rm AH}$ range from zero to unity in increments of $0.1$.   
The graph includes eleven solid lines (representing an evaluation of  
the binding energy (\ref{Mbind}) in maximal slicing) and eleven  
dashed lines (for Kerr-Schild slicing).  All 22 lines coincide within  
our numerical error.  We computed these sequences using $N = 26$  
collocation points in each domain.}
\label{largeRatio}
\end{figure}

We now construct constant-mass sequences, meaning sequences of  
varying shell radius $r_{\rm SH}$ but constant shell rest mass $M_ 
{\rm SH}$ and black hole irreducible mass $M_{\rm BH}$.   
Specifically, we focus on ``extreme-mass-ratio" sequences with $M_ 
{BH}/M_{\rm SH} = 1000$ and equal-mass sequences $M_{\rm BH}/M_{\rm  
SH}=1$.  For both choices of the mass ratio we construct sequences  
for our two choices of the extrinsic curvature (\ref{K}), maximal  
slicing and Kerr-Schild slicing, and for the horizon lapse (\ref 
{alphaHorBound}) ranging from $\alpha_{\rm AH} = 0$ to $\alpha_{\rm  
AH} = 1$ in increments of 0.1.

In the following, we will graph the binding energy (\ref{Mbind}) as a  
function of the areal radius.  Typically, the binding energy $M_{\rm  
B}$ is only a small fraction of the involved masses, meaning that the  
relative error in the binding energy is larger than that for the  
masses -- and hence the fields -- themselves.   We found that, to  
achieve similar accuracy, the extreme-mass-ratio sequences (for which  
the binding energy is much smaller than for the equal-mass  
sequences)  required slightly more collocation points than the 
equal-mass sequences.

In Fig.~\ref{largeRatio} we show the binding energy for extreme-mass- 
ratio sequences with $M_{\rm BH}/M_{\rm SH} = 1000$.   The graph  
represents 22 plots, corresponding to the eleven different values of $ 
\alpha_{\rm AH}$ and to evaluating the ADM binding energy (\ref 
{Mbind}) for both maximal slicing and Kerr-Schild slicing.  To within the 
accuracy of our code, all 22 lines agree with each other, so that they all lie
on top of each other and appear as one line in Fig.~\ref{largeRatio}.

The minimum of the binding energy corresponds to the innermost stable  
circular orbit (ISCO).  In the extreme-mass-ratio limit we may  
neglect the particles' self-gravity, so that we are effectively  
solving for a test-particle in the Schwarzschild geometry.  As  
expected, we find that the ISCO is located at $R=6M_{\rm BH}$,  
representing another independent test of our code.

\begin{figure}[t]
\includegraphics[width=3in]{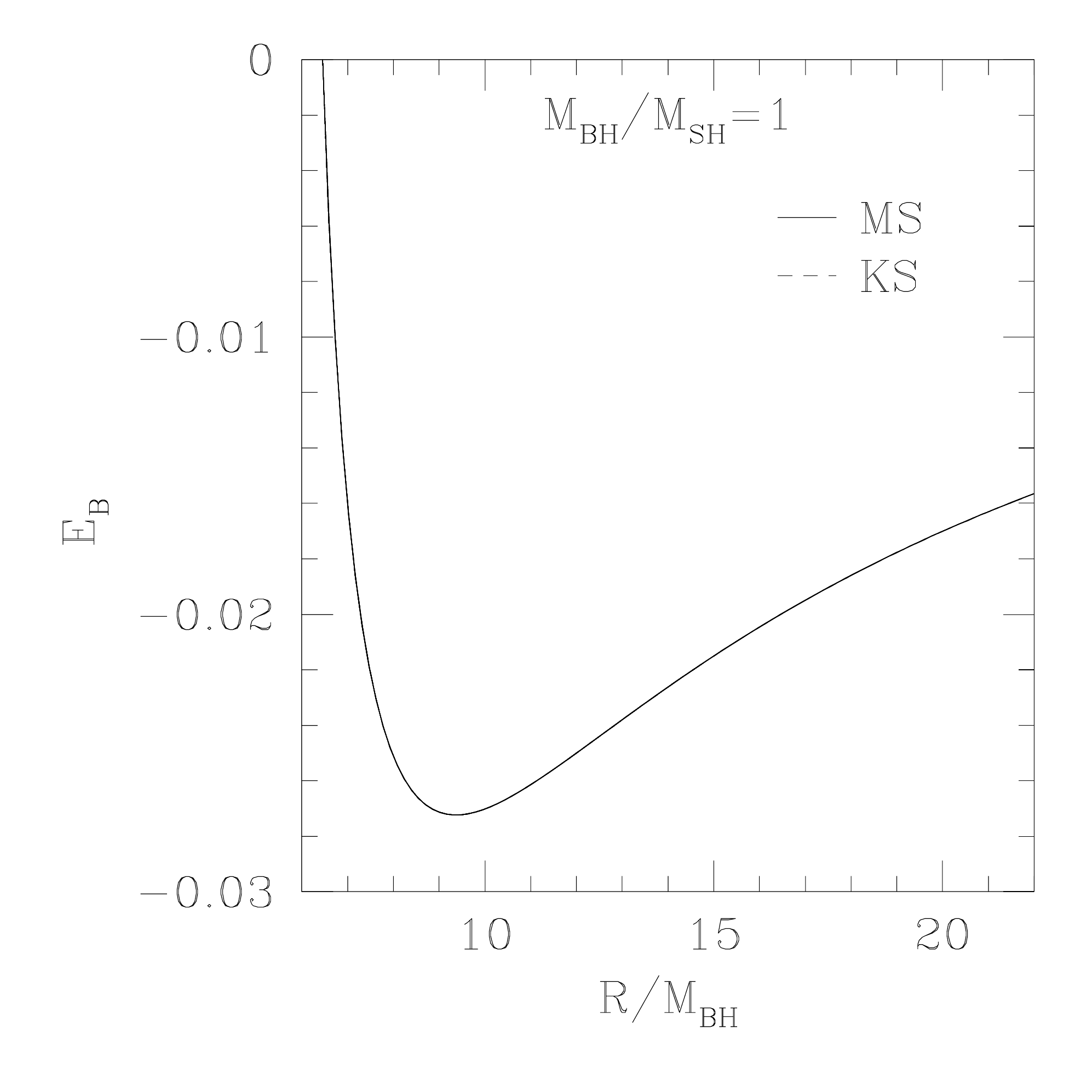}
\caption{The ADM binding energy $E_{\rm B}$ as a function of the  
shell's areal radius $R$ for an equal mass sequence, $M_{BH}/M_{SH} =  
1$.  As in Fig.~\ref{largeRatio}, values of $\alpha_{\rm AH}$ again  
range from zero to unity in increments of $0.1$, and the graph again  
contains eleven lines each for Maximal and Kerr-Schild slicing.  All  
22 lines again coincide within our numerical error.  We obtained  
these results with $N=20$ collocation points in each domain.}
\label{equalRatio}
\end{figure}

In Fig.~\ref{equalRatio} we show the equivalent graphs for equal-mass  
sequences with $M_{\rm BH}/M_{\rm SH} = 1$.   As for the extreme-mass- 
ratio sequences in Fig.~\ref{largeRatio} all 22 graphs coincide to  
within our numerical accuracy.   We note that the ISCO is now located  
at a larger radius of about $9.367 M_{\rm BH}$.

Even in this case, we do not find any evidence that the choice of the  
slicing condition (\ref{K}), or the choice of the boundary condition  
for the lapse on the horizon (\ref{alphaHorBound}), have any effect  
on the physical content of our solutions.  Clearly, these different  
choices lead to different solutions for the conformal factor $\psi$,  
the lapse $\alpha$ and the shift $\beta$.   When expressed in terms  
of gauge-invariant quantities, however, all our solutions become  
indistinguishable to within the accuracy of our numerical code.

\section{Summary}

We solve Einstein's constraint equations in the extended conformal  
thin-sandwich decomposition to construct spherical shells of 
non-interacting, isotropic particles in circular orbit about a 
non-rotating black hole.  We construct these solutions for both maximal  
slicing and Kerr-Schild slicing (see equations (\ref{K})), and for a  
number of different choices for the horizon lapse.   These different  
choices lead to very different solutions for the lapse $\alpha$, the  
shift $\beta^i$ and the conformal factor $\psi$.  However, when  
expressed in terms of gauge-invariant quantities -- for example the  
binding energy as a function of the shell's areal radius for given  
shell and black hole masses -- our solutions become indistinguishable.
At least in the limited context of our spherically symmetric  
solutions, these findings provide no evidence that the choices for  
the mean curvature and horizon lapse affect the physical content of  
the solutions.

\acknowledgments

KM gratefully acknowledges support from the Coles Undergraduate  
Research Fellowship Fund at Bowdoin College and from the Maine Space  
Grant Consortium.  KM and TWB would like to thank the Laboratoire  
Univers et Theories at the Observatoire de Paris, Meudon, for its  
hospitality.  This work was supported in part by NSF grant  
PHY-0456917 to Bowdoin College and
by the ANR grant 06-2-134423 \emph{M\'ethodes math\'ematiques pour la  
relativit\'e g\'en\'erale}.
\begin{appendix}

\section{An analytical solution}
\label{appA}

For maximal slicing ($K = 0$) and the lapse boundary condition $ 
\alpha_{\rm AH} = 0$, we can find an analytic solution to the  
differential equations (\ref{DiffEq}), subject to the boundary  
conditions (\ref{infBound}) and (\ref{InBoundCond}) as well as the  
jump conditions (\ref{jumpConditions}).
The solution depends on the input parameters $M_{\rm SH}$, $W$, $r_ 
{\rm BH}$, and $r_{\rm SH}$.   We may then enforce the geodesic  
condition (\ref{geodesicCondition}) to eliminate one
of these variables, and compute the mass of the black hole from (\ref 
{Mirr}).  The solution presented here represents a generalization of  
the solutions of \cite{SkoB02}, who considered the same system but  
without a black hole.

We begin with the momentum constraint (\ref{betaConstraintMod}).  For  
$K = 0$ and $\alpha_{\rm AH} = 0$ we find that
\begin{equation} \label{analyticBetaSH}
\beta = 0
\end{equation}
is a self-consistent solution to both the equation and its boundary  
conditions.
This implies that Eqs.~(\ref{psiConstraintMod}) and (\ref 
{alphaConstraintMod})
for respectively the conformal factor $\psi$ and the combination $ 
\alpha \psi$
reduce to flat Laplace equations in the vacuum regions away from the  
shell.  In spherical symmetry, the only possible solutions are of the  
form $k_1 + k_2/r$, where $k_1$ and $k_2$ are arbitrary constants  
that have to be determined from the boundary conditions.  For each  
function we need four conditions to determine these constants both in  
the interior and the exterior of the shell.  These four conditions  
arise from the
outer boundary conditions (\ref{infBound}), the inner boundary  
conditions (\ref{InBoundCond}), continuity of the functions at the  
shell, and the jump conditions (\ref{jumpConditions}) for their first  
derivatives.   Using these conditions, we find
\begin{equation} \label{analyticPsiSH}
\psi = \left\{
\begin{array}{ll} a \left(1+\dfrac{r_{\rm BH}}{r}\right) \, , & r  
\leq r_{\rm SH} \\[3mm]
      1+ \dfrac{a \left(r_{\rm SH}+r_{\rm BH} \right) - r_{\rm SH} } 
{r} \, ,&   r \geq r_{\rm SH}
\end{array}
\right.
\end{equation}
for the conformal factor and
\begin{equation} \label{analyticAlphaSH}
\alpha = \left\{
\begin{array}{ll}
\displaystyle \dfrac{c \left(1-r_{\rm BH}/r \right)}{a \left(1+r_{\rm  
SH}/r \right)} \, , & r < r_{\rm SH} \\[3mm]
\displaystyle \dfrac{1+\left( c \left( r_{\rm BH} - r_{\rm SH}  
\right) - r_{\rm SH} \right)/r}{1+\left(a \left(r_{\rm SH}
+r_{\rm BH} \right)-r_{\rm SH} \right)/r } \, , & r \geq r_{\rm SH}
\end{array}
\right.
\end{equation}
for the lapse.  Here the constants $a$ and $c$ are given by
\begin{equation}\label{aExpression}
a = 1 + \frac{M_{\rm SH}W}{2r_{\rm SH} \psi_{\rm SH}}
\end{equation}
and
\begin{equation}\label{cExpression}
c = 1 - \frac{M_{\rm SH} \alpha_{\rm SH} \left( 3 W^2 -2 \right)}{r_ 
{\rm SH} \left( 2 W \psi_{\rm SH} \right)}.
\end{equation}
Inserting these constants, which themselves depend on the values of  
the conformal factor and the lapse at the shell, into (\ref 
{analyticPsiSH}) and (\ref{analyticAlphaSH}), we find
\begin{equation}\label{psiHorizon}
\psi_{\rm SH} = \frac{1}{2} \left(1+\frac{r_{\rm BH}}{r_{\rm SH}}  
\right) \left[ 1+ \left( 1+ \frac{2 M_{\rm SH} W}{r_{\rm BH}+r_{\rm  
SH}} \right)^{1/2} \right]
\end{equation}
and
\begin{equation}\label{alphaHorizon}
\alpha_{\rm SH} = \left( p + \frac{\left(3+p\right)W^2-2}{2r_{\rm SH} 
\psi_{\rm SH}W} \right)^{-1}.
\end{equation}
Here we have abbreviated $p\equiv(r_{\rm SH}+r_{\rm BH})/(r_{\rm SH}- 
r_{\rm BH})$.

So far these solutions depend on all four parameters $M_{\rm SH}$, $W 
$, $r_{\rm BH}$, and $r_{\rm SH}$.   We now find a relation between  
these parameters by inserting (\ref{analyticPsiSH}) and (\ref 
{analyticAlphaSH}) into the geodesic equation (\ref{geodesicMod}),  
which yields
\begin{eqnarray}\label{omegaAnalytic}
&& \Omega^2 = \left(\dfrac{u^{\phi}}{u^t}\right)^2 = -\dfrac{^{(4)} 
\Gamma^r_{tt}}{^{(4)}\Gamma^r_{\phi\phi}} \\
&& = \dfrac{r_{\rm SH}^3 c \left(r_{\rm BH} - r_{\rm SH}\right)\left 
(r_{\rm SH}\left(a-c\right)+r_{\rm BH}\left(a+c+2ac\right)\right)}
          {2\left(a\left(r_{\rm SH}+r_{\rm BH}\right)\right)^6\left 
(ar_{\rm BH}-r_{\rm SH}\right)}. \nonumber
\end{eqnarray}
Unfortunately, this expression is not very useful for our purposes in  
this form.  We find an alternative form by evaluating (\ref 
{geodesicCondition}) for our solution (\ref{analyticPsiSH}) and (\ref 
{analyticAlphaSH}),
which results in a fifth order polynomial for $W$,
\begin{eqnarray}\label{geoAnalytic}
0 &=& 4(r_{\rm BH}-r_{\rm SH})^3 r_{\rm SH} \nonumber \\
&& + M_{\rm SH} (r_{\rm BH}-r_{\rm SH})^2(r_{\rm BH}+r_{\rm SH})W  
\nonumber \\
&&- 2r_{\rm SH}(r_{\rm BH}-r_{\rm SH})(5r_{\rm BH}^2-14r_{\rm BH}r_ 
{\rm SH}+5r_{\rm SH}^2)W^2 \nonumber \\
&&- 2M_{\rm SH}(r_{\rm BH}-2r_{\rm SH})(r_{\rm BH}-r_{\rm SH})(r_{\rm  
BH}+r_{\rm SH})W^3 \nonumber \\
&&+ 6r_{\rm SH}(r_{\rm BH}-r_{\rm SH})(r_{\rm BH}^2-4r_{\rm BH}r_{\rm  
SH}+r_{\rm SH}^2)W^4 \nonumber \\
&&+ M_{\rm SH} (r_{\rm BH}-2r_{\rm SH})^2(r_{\rm BH}+r_{\rm SH})W^5
\end{eqnarray}
In the limit of a zero-mass black hole we have $r_{\rm BH} = 0$ and  
(\ref{geoAnalytic}) reduces to
\begin{eqnarray}
0&=&-4r_{\rm SH} + M_{\rm SH} W + 10 r_{\rm SH} W^2 - 4 M_{\rm SH}  
W^3 \nonumber \\
&&- 6 r_{\rm SH} W^4 + 4 M_{\rm SH} W^5,
\end{eqnarray}
which agrees with the corresponding equation (26) of \cite{SkoB02}.

Instead of parameterizing the solution by $r_{\rm BH}$, it is more  
desirable to fix the black hole mass $M_{\rm BH} = M_{\rm irr}$.   
Towards this end we combine (\ref{geoAnalytic}) with (\ref 
{analyticPsiSH}) and (\ref{Mirr}), which results in a quadratic  
equation for $r_{\rm SH}$.  The solutions to this equation are
\begin{eqnarray}\label{constAnalytic}
r_{\rm SH} &=& \Big[4M_{\rm BH}r_{\rm BH}^2 + 2 M_{\rm BH} M_{\rm SH}  
r_{\rm BH} W - 2 M_{\rm BH}^2 r_{\rm BH} \nonumber \\
&&\pm  2\sqrt{2} \big(M_{\rm BH} M_{\rm SH}^2 r_{\rm BH}^3 W^2\big)^ 
{1/2} \Big]/ \nonumber \\
&& \Big[ 2(M_{\rm BH}^2-2M_{\rm BH} r_{\rm BH}) \Big] .
\end{eqnarray}
We henceforth ignore the ``$-$" solution, as only the ``$+$" solution  
refers to stable solutions.

Finally, we may insert (\ref{constAnalytic}) back into (\ref 
{geoAnalytic}), which results into a polynomial for $r_{\rm BH}$ that  
can be factored into two cubic polynomials
\begin{eqnarray} \label{cubicPol1}
0 & = & \left( a_3 \bar r_{\rm BH}^3 + a_2 \bar r_{\rm BH}^2 + a_1  
\bar r_{\rm BH} + a_0 \right) \times
\nonumber \\
& & \left( b_3 \bar r_{\rm BH}^3 + b_2 \bar r_{\rm BH}^2 + b_1 \bar r_ 
{\rm BH} + b_0 \right).
\end{eqnarray}
Here we have abbreviated $\bar r_{\rm BH} \equiv r_{\rm BH}/M_{\rm SH} 
$, and the coefficients $a_i$ and $b_i$ are, in terms of the mass ratio $q \equiv M_ 
{\rm BH}/M_{\rm SH}$,
\begin{eqnarray}
a_3 & = & b_3 = 32 q^2(-2+3W^2)^2 \\
a_2 & = & -16 q(-2+3W^2)[-4 q^2 + 6 q W \nonumber \\
&& + (6q^2-1)W^2 -9qW^3 +2W^4] \nonumber \\
b_2 & = & -4q(-2+3W^2)^2(8q^2-8qW+5W^2) \nonumber \\
a_1 & = & 2[16 q^4 - 48 q^3 W + (44 q^2 - 48 q^4) W^2 \nonumber \\
&& +(144 q^3-12 q) W^3+(36q^4-136q^2 \nonumber \\
&& +1)W^4 +(38 q-108 q^3)W^5+(105 q^2 \nonumber \\
&& - 2) W^6 -30qW^7+W^8] \nonumber \\
b_1 & = & 2(-2+3W^2)[-8q^4+16q^3W \nonumber \\
&& +(12q^4-20q^2) W^2 +(10q-24q^3)W^3 \nonumber \\
&& +(26q^2-2) W^4-12qW^5+3W^6] \nonumber \\
a_0 & = & b_0 = -qW^2(-2q+W+3qW^2-2W^3)^2  \nonumber
\end{eqnarray}

We can now construct a solution for given masses $M_{\rm BH}$ and $M_ 
{\rm SH}$ and Lorentz factor $W$ as follows.  Given the mass ratio $q 
$ and $W$ we first find the six solutions for $\bar r_{\rm BH}$ from  
the two polynomials in (\ref{cubicPol1}).  We then insert the  
corresponding values $r_{\rm BH}$ into (\ref{constAnalytic}), which  
yields six solutions for $r_{\rm SH}$.  The largest real solution is  
the solution of interest; we keep only this solution as well as its  
corresponding value of $r_{\rm BH}$ and disregard all others.   These  
values can then be inserted into (\ref{analyticPsiSH}) and (\ref 
{analyticAlphaSH}), which determines the solution completely.


\end{appendix}

\end{document}